\documentstyle[11pt,aaspp4]{article}

\begin{document}
\title{NATURE OF HARD X-RAY SOURCE FROM OPTICAL IDENTIFICATION \\
OF THE ASCA LARGE SKY SURVEY}
\author{Masayuki Akiyama}
\affil{SUBARU Telescope NAOJ, 650 North A'ohoku Place, Hilo, 96720, HI, U.S.A \\
/ Department of Astronomy, Kyoto University}
\author{Kouji Ohta}
\affil{Department of Astronomy, Kyoto University}
\author{Toru Yamada}
\affil{Astronomical Institute, Tohoku University}
\author{Yoshihiro Ueda and Tadayuki Takahashi}
\affil{ISAS}
\author{Masaaki Sakano and Takeshi Tsuru}
\affil{Department of Physics, Kyoto University}
\author{Ingo Lehmann and G\"unther Hasinger}
\affil{Astrophysikalisches Institut Potsdam}

\begin{abstract}

We present results of optical identification of the ASCA Large Sky Survey.
X-ray sources which have hard X-ray spectra were identified with
type-2 AGN at redshifts smaller than 0.5. It is supported that the 
absorbed X-ray spectrum of type-2 AGN makes
the Cosmic X-ray Background harder in the hard X-ray band
than type-1 AGN, which is  
main contributer in the soft X-ray band.
Absence of type-2 AGN at redshift larger than 1 in the identified
sample, which contrasts to the existence of 6 broad-line QSOs,
suggests a deficiency of so-called ``type-2 QSO'' at high redshift.

\end{abstract}

\section{Introduction}

In order to understand the origin of the Cosmic X-ray Background (CXB),
especially to reveal the nature of the hard X-ray sources, which are
required to account for the spectrum of the CXB,
we are making a wide area X-ray source survey using the ASCA
satellite in the 0.7--10 keV band (the ASCA Large Sky Survey hereafter LSS,
\cite{ued98a}, \cite{ued98b}, \cite{ued98c}). 
It covered 7 square degree of the sky near the north galactic pole 
down to the X-ray flux of 10$^{-13}$ erg/sec/cm$^2$
in the 2--10 keV band.
We have detected 44 sources in the 2--10 keV band 
above 3.5$\sigma$ by the GIS and SIS and resolved 23\%
of the CXB. Significant fraction of detected X-ray sources had a 
hard photon index (smaller than 1.0) and the averaged X-ray spectrum
of the sources have a photon index of 1.49$\pm$0.10 in the 2--10 keV band 
(\cite{ued98a});
closer to that of the CXB ($\Gamma \sim$1.4: e.g., \cite{gen95}) than that of type-1 AGNs (
$\Gamma \sim$1.7: e.g., \cite{tur89}),
which are the main contributer to the CXB in the 0.5--2 keV ROSAT band
(e.g., \cite{sch98}).
A good candidate for these hard sources is a 
type-2 AGN (e.g., \cite{com95}).
In this presentation, we summarize the results of optical identification
of the LSS.
Throughout this paper, we used H$_0$=50km/sec/Mpc and q$_0$=0.5.

\section{Optical Follow-up Observation of the LSS and reliability of 
the identification}

We have made observations for optical identification of 34 X-ray sources
detected above 3.5$\sigma$ with the SIS in the 2--7 keV band.
We selected candidates of optical counterparts using
CCD images taken with the KISO Schmidt telescope and the APM catalog (\cite{mcm92}).
For some sources, deep images were obtained with the 
University of Hawaii $88^{\prime\prime}$ telescope.
An error radius of an ASCA source is estimated to be 0.6--0.8 arcmin
(\cite{ued98c}) 
and there are some optical objects within each error circle.
We have selected targets for spectroscopy using various catalogs and
optical color and magnitude as shown below.
\begin{enumerate}
\vspace{-0.5mm}
\item{ROSAT data :
We made ROSAT HRI observations of the selected areas (2 fields of HRI).
Almost all LSS sources in the field were detected by the HRI and
their optical counterparts were pinpointed.
ROSAT PSPC source catalog (\cite{vog98}) was also used for other sources.}

\item{VLA FIRST catalog (\cite{whi97}) :
About half (13/34) of the sample were 
detected in the FIRST survey and optical counterpart was pinpointed
thanks to the good positional accuracy of the FIRST catalog.
The high fraction of radio-detected source is due to nice match
between the limit ($\sim$ 1mJy) of VLA FIRST 1.4 GHz survey 
and that of the LSS; ratio of the flux limits 
is comparable to 
the radio to X-ray flux ratio of {\it radio-quiet} type-1 
AGN (e.g., \cite{elv94}). 
It is worth noting that radio emission is transparent against
obscuring material, thus obscured AGNs are unbiasly
detected in the radio wavelength as well as hard X-ray.
Chance coincidence of a radio source which is not an X-ray counterpart
is expected to be 0.5 for the whole sample and is negligible.}

\item{Optical magnitude, color and galaxy excess :
As a type-1 AGN candidate, we selected optical objects using optical
magnitude and color (\cite{aki98a}).
X-ray sources associated with an excess of galaxies
in the optical image were identified
with clusters of galaxies.}

\end{enumerate}
Most sources have at least one object which meets one of the above
criteria. We have made spectroscopic observations
for these objects with the highest priority. 

Spectroscopic observations were made at the UH 88$^{\prime\prime}$ telescope
at Mauna Kea. 
The X-ray sources were identified with
19 type-1 AGNs, 3 type-1.5 AGNs which show a broad H$\alpha$
but no broad H$\beta$ line,
6 type-2 AGNs which show AGN-like line ratios and no broad line, 
2 clusters of galaxies (not spectroscopically confirmed yet), and one galactic star. 3 sources are
still unidentified.

We have checked the reliability of our identification
by estimating chance contaminations.
Considering number counts of type-1 AGN (\cite{har90}),
the expected number of chance contaminations of type-1 AGN 
brighter than B=21.0 with redshift smaller than 3 
within 34 error circles is 0.29 and negligible.
For type-2 AGNs, 4/6 of them are pinpointed by FIRST radio survey and 
chance contamination is very small for the sample as mentioned above.
One of the remaining type-2 AGNs is pointed out by ROSAT PSPC. 
Therefore our identifications are reliable.

\begin{figure}
  \epsscale{1.0}
  \begin{center}
    \plotfiddle{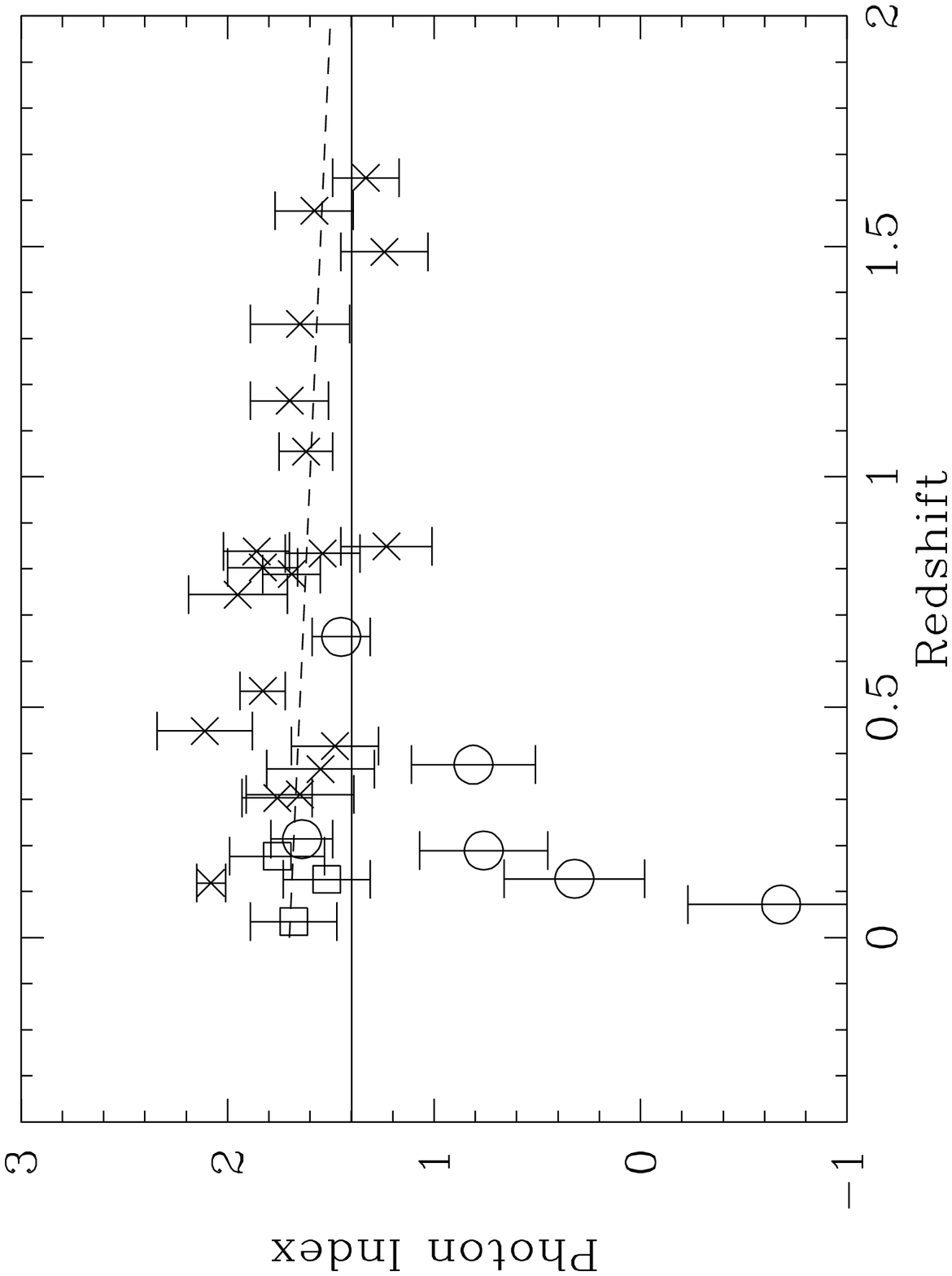}{50pt}{270}{37}{37}{-270}{130}
    \plotfiddle{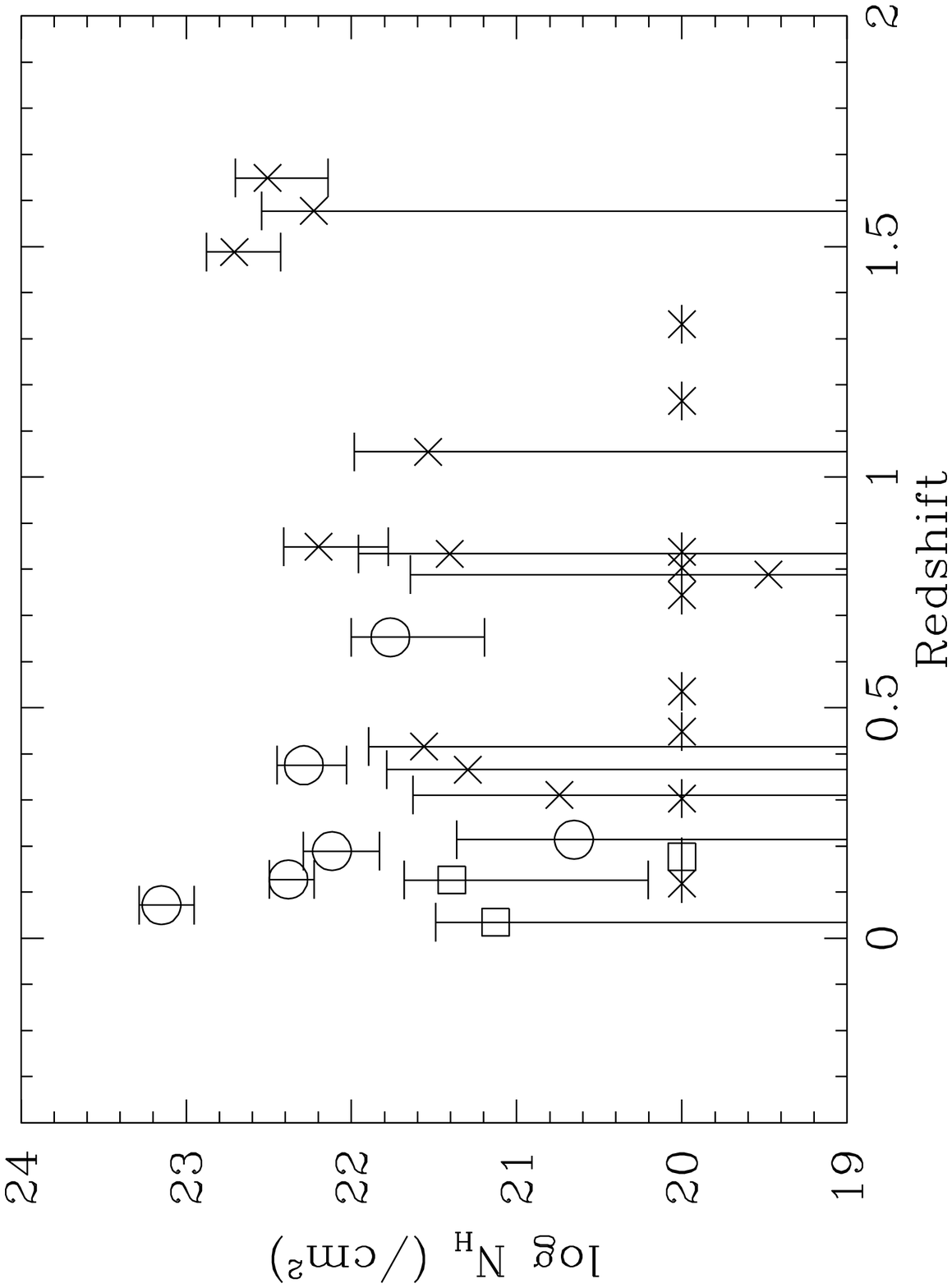}{50pt}{270}{37}{37}{-10}{195}
  \end{center}
  \caption{Left: Photon index versus redshift of the identified AGNs.
Cross, rectangle, and circle represent type-1 AGN, type-1.5 AGN, 
and type-2 AGN, respectively.  
Horizontal solid line is the photon index of the CXB.
Expected change of apparent photon index (0.7--10 keV) for a 
type-1 Seyfert with redshift is shown as a dashed line. 
It is derived by using the average X-ray 
spectrum of nearby type-1 Seyferts (\cite{gon96}).
Right: Column density versus redshift. Marks are the same as in the left panel.
Fitting is made by assuming  an intrinsic photon index of 1.7
and an intrinsic absorption at the object redshift.
X-ray sources which have photon index larger than 1.7 is 
plotted at column density of 10$^{20}$/cm$^2$.
Photon index and column density are determined in the 0.7--10 keV
band. \label{figA}}
\end{figure}

\begin{figure}
  \epsscale{1.0}
  \begin{center}
    \plotfiddle{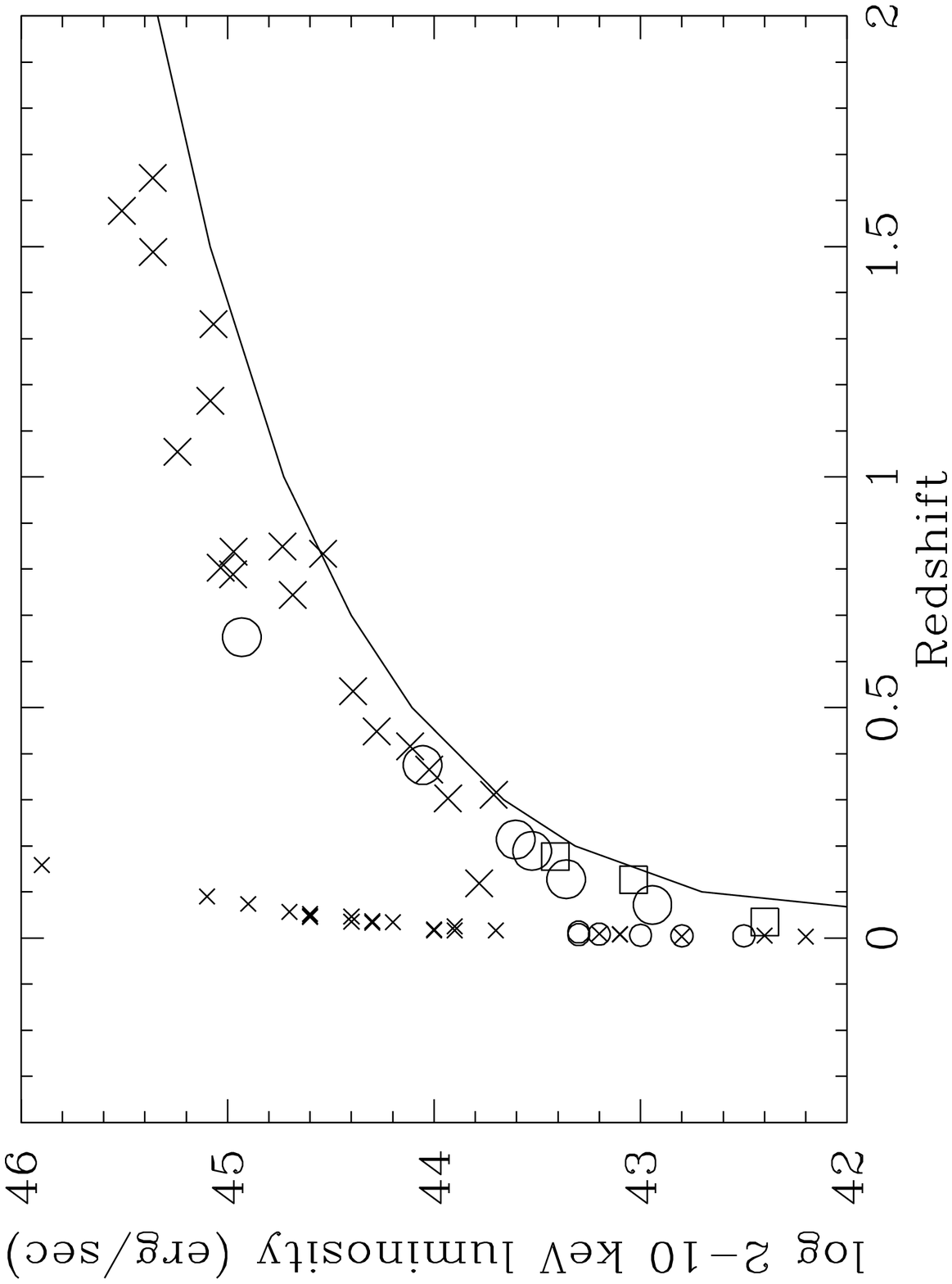}{50pt}{270}{37}{37}{-270}{130}
    \plotfiddle{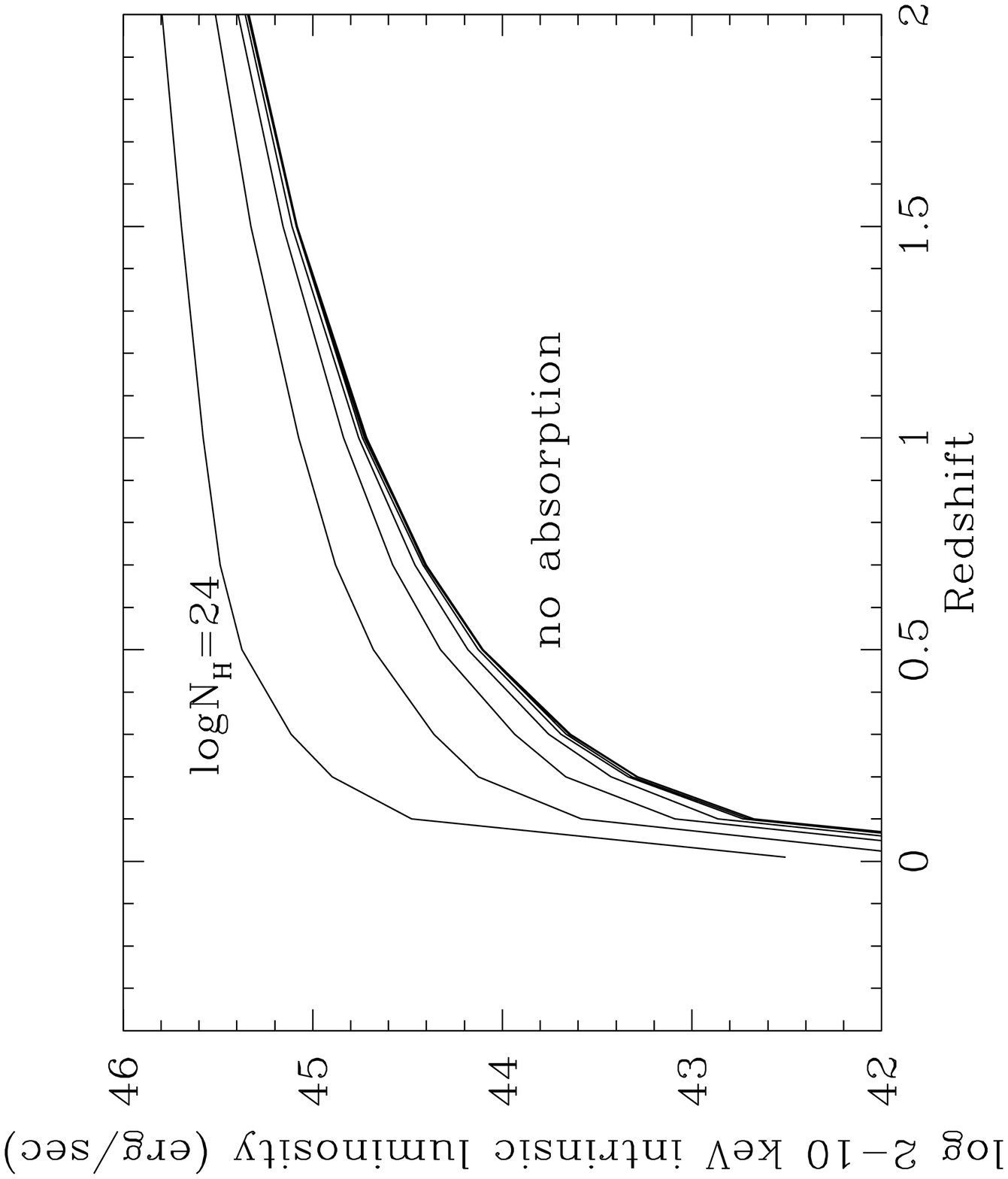}{50pt}{270}{37}{37}{-10}{195}
  \end{center}
  \caption{Left: 2--10 keV luminosity versus redshift. The HEAO-1 A2
AGN sample (\cite{tur89}) is also shown with small marks. For the LSS sample,
the luminosity is not corrected for intrinsic absorption.
Marks are the same as in figure 1. Solid line is the detection limit
for object without intrinsic absorption.
 Right: Detection limit to the intrinsic 2--10 keV luminosity 
as a function of redshift. 
From top to bottom, intrinsic column density
of log N$_{\rm H}$ = 24.0, 23.5, 23.0, 22.5, 22.0, 21.5, 21.0, and 
no absorption with an intrinsic photon index of 1.7 were assumed.
The survey limit of the SIS 3.5$\sigma$ is 1.2 cts/ksec in the 2--7
keV band. \label{figC}}
\end{figure}

\begin{figure}[t]
  \epsscale{1.0}
  \begin{center}    
    \plotfiddle{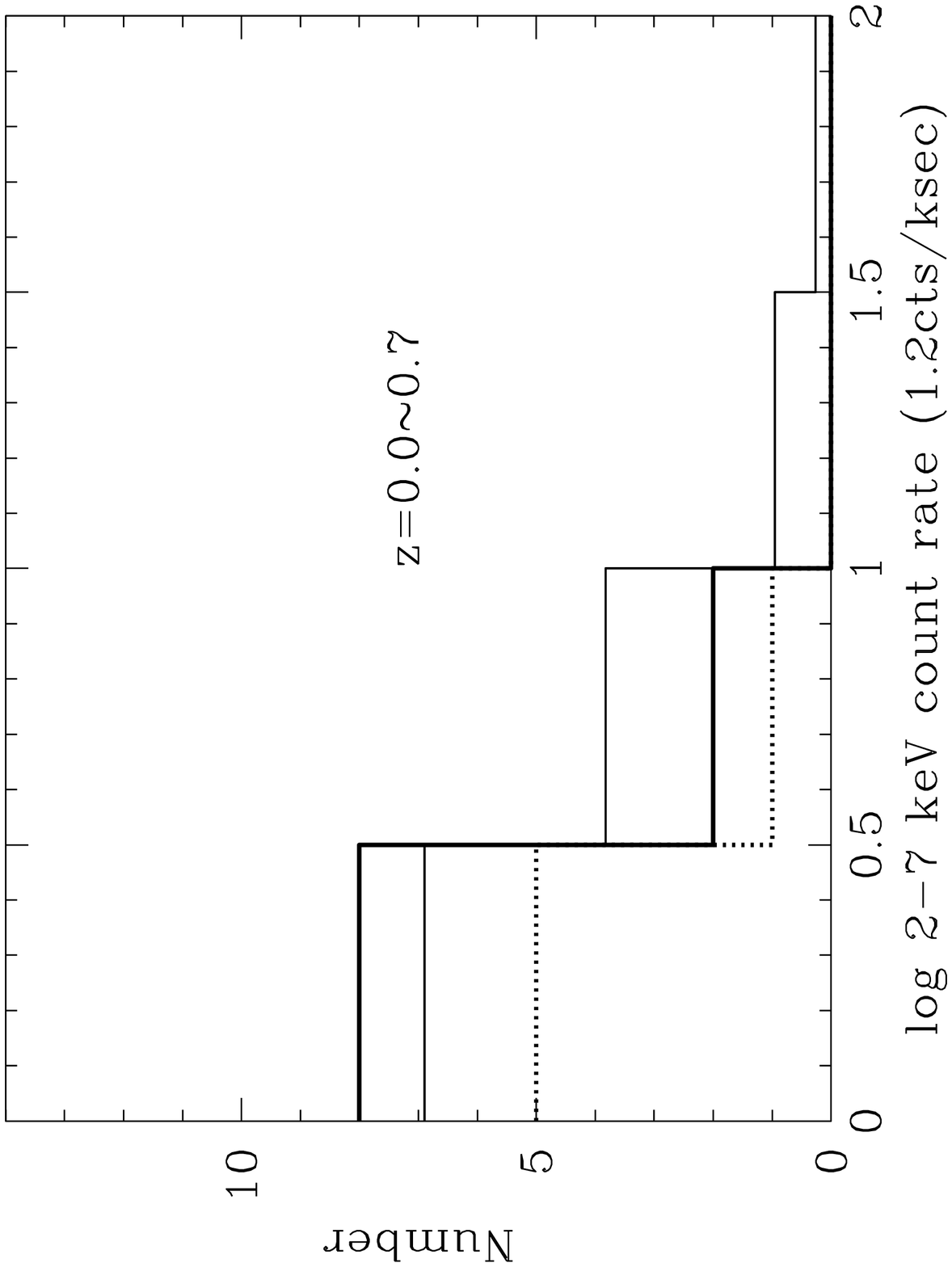}{40pt}{270}{35}{35}{-260}{130}
    \plotfiddle{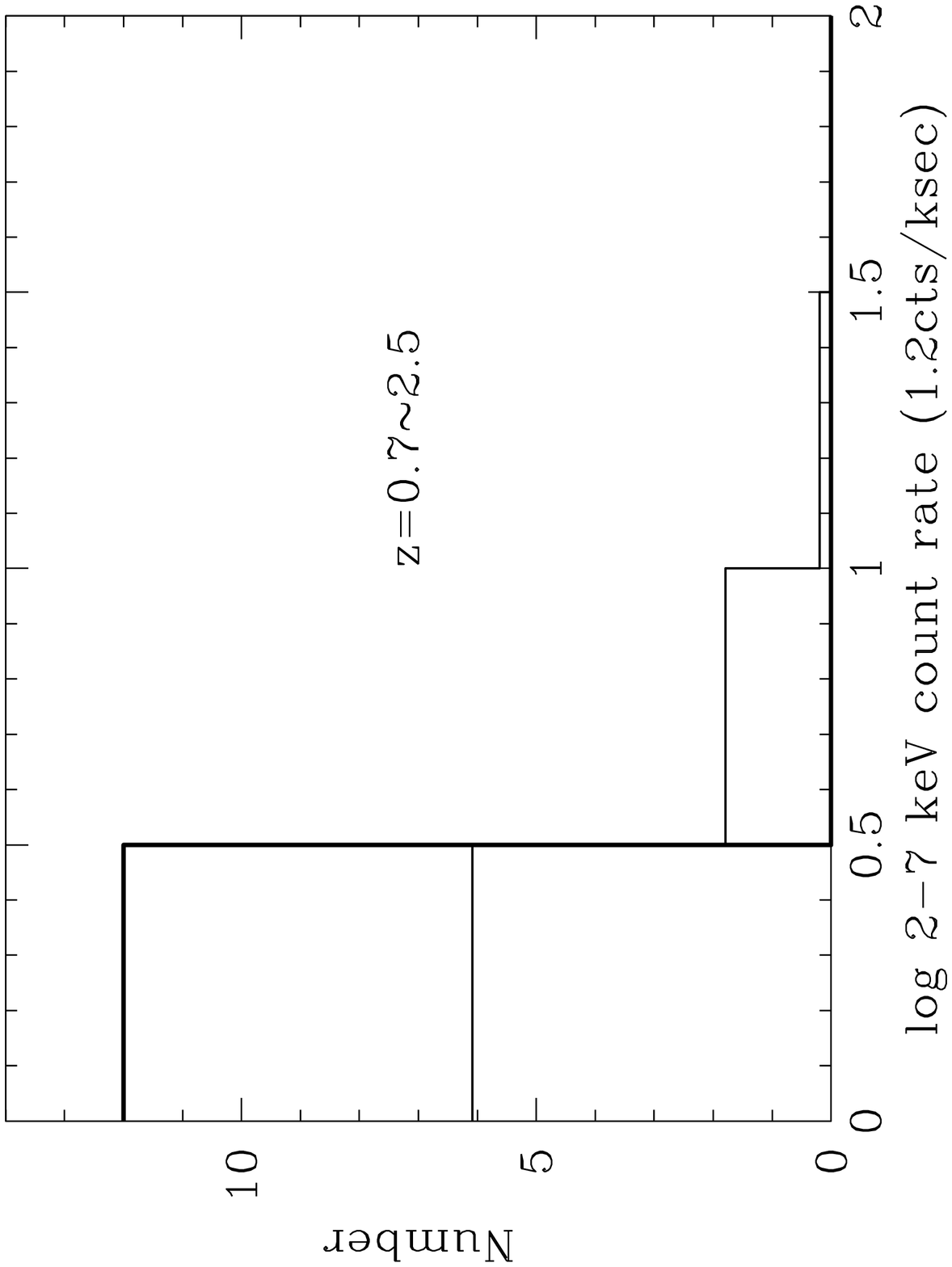}{40pt}{270}{35}{35}{-20}{185}
  \end{center}
  \caption{Detected number of AGN as a function of SIS count rate.
The identified sample is divided at redshift of 0.7.
Left: low redshift sample. Right: high redshift sample.
Thick solid line represents type-1 plus type-1.5 AGNs and
dotted line shows type-2 AGNs.
Thin solid line shows expected number of type-1 AGNs derived from
X-ray luminosity function of AGNs in the soft band (see section 4).
\label{figD}}
\end{figure}

\section{Nature of the Hard Sources}

The distribution of the observed photon index as a function of redshift is
shown in figure \ref{figA}. 
All X-ray sources which have photon index smaller than 1
are identified with type-2 AGNs at redshifts smaller than 0.5.
For example, the hardest source was identified with a type-2 Seyfert
at z=0.072 (\cite{aki98b}).
A dashed line in figure \ref{figA}
shows an expected change of {\it apparent} photon index
(0.7--10 keV) for a type-1 Seyfert with redshift.
Because of the reflection component,
observed photon index of a type-1 Seyfert in the 0.7--10 keV band
is expected to be getting harder to z$\sim$2.
The photon index distribution of the type-1 AGN sample is
consistent with this decreasing trend.
It should be noted that at larger redshift the apparent 
photon index turns to larger and will not be as hard as the CXB.

We have derived column densities of absorbing matter at the
object's redshift assuming an intrinsic photon index of 1.7
(figure \ref{figA} right). 
The type-2 AGNs which have hard X-ray spectra are fitted with
column density of 10$^{22}\sim$10$^{23}$ /cm$^{2}$. 
The upper boundary of the distribution of the column density (N$_{\rm H}\sim 10^{22.5}$ 
/cm$^2$) is consistent with an estimated limit on the column
density in  the LSS sample (see figure \ref{figC}).
Three high-redshift type-1 AGNs are fitted with large column densities.
This can be explained by the effect of the reflection component as 
shown above.

These results support an idea that the absorbed X-ray spectra 
of type-2 AGNs make the CXB harder than type-1 AGN (\cite{com95}).
However there is no hard X-ray source which identified with
object at high redshift. This seems to be inconsistent with 
the CXB model which assume that the type-1 to type-2 AGN ratio does not
change with redshift and luminosity.
In the next section we examine the redshift and luminosity distribution
of the identified AGNs.

\section{Deficiency of type-2 QSOs ?}

X-ray luminosity vs. redshift of identified AGNs is shown in figure 
\ref{figC}.
(We also plotted AGNs in the HEAO-1 A2 sample (\cite{tur89})
with small marks for comparison.)
The number ratio between type-1 plus type-1.5 AGNs to type-2 AGNs
is clearly changing with redshift, luminosity or both.
In the next figure, detection limits to the intrinsic luminosity of
objects with various absorption column density are shown as a function 
of redshift, for the SIS 3.5$\sigma$ sample.
We can unbiasly detect objects 
with intrinsic column density of up to $10^{22.5}$/cm$^2$ in redshift 
smaller than 1.5 and up to $10^{23}$/cm$^2$ in redshift larger than 1.5,
thanks to the redshift effect.
Thus, if the column density distribution of absorbing matter  
and the critical column density which divides type-1 and type-2 AGN 
changes with neither redshift nor luminosity, we expect the number ratio of type-2 to
type-1 AGN does not change or become larger at higher redshift in our sample. 
The smaller number ratio at higher redshift
supports the deficiency (e.g, \cite{law91}) 
of so called ``type-2 QSO'', which is
a narrow-line luminous AGN.
It should be noted that
there remains three sources which are unidentified in the SIS 3.5$\sigma$ sample
and these objects could be the missing high redshift type-2 AGNs.

To compare the redshift distribution of the identified AGNs
with an AGN luminosity function and its evolution derived by
ROSAT surveys, we divide the sample at redshift of 0.7.
In figure \ref{figD}, we show histograms of the detected number of type-1 plus type-1.5
and type-2 AGN as a function of SIS count rate. 
Expected numbers of AGNs using an AGN luminosity function 
derived from various ROSAT surveys (\cite{has98}) is also shown.
Since most of the ROSAT sample is type-1 AGN,
we assumed the canonical photon index of 1.7 for type-1 AGNs
to convert the 0.5--2 keV luminosity to SIS count rate.
Detected number of type-1 plus 
type-1.5 AGN in the redshift range of 0.0--0.7 is 
consistent with the estimated number. 
While in redshift range of 0.7--2.5,
the detected number shows a small excess.
This could be explained by an excess of
QSOs at z=0.8 in our sample (see, figure \ref{figC}).
The number of type-2 AGNs in the low redshift sample is about half of
the number of type-1 AGNs. 
On the other hand, there is no type-2 AGN in the high redshift
sample as mentioned above. 

Compilation of optical identifications of deep ASCA surveys will 
set more strict constraints on the change of the number ratio of 
type-2 to type-1 AGN with redshift and luminosity. 
However, XMM surveys, which cover flux limit range from comparable
to to 100 times deeper than that of the LSS, are absolutely needed
to qualitatively discuss cosmological evolutions of the luminosity
functions of type-2 and type-1 AGN.


\begin{thebibliography}{}

\bibitem[Akiyama et al. 1998a]{aki98a}
Akiyama, M., et al. 1998, AN, 319, 63

\bibitem[Akiyama et al. 1998b]{aki98b}
Akiyama, M., et al. 1998, \apj, 500, 173

\bibitem[Comastri et al. 1995]{com95}
Comastri, A., Setti, G., Zamorani, G., \& Hasinger, G. 1995, A\&A, 296, 1

\bibitem[Elvis et al. 1994]{elv94}
Elvis, M., et al.,  1994, \apjs, 95, 1 

\bibitem[Gendreau et al. 1995]{gen95}
Gendreau, K.C., et al. 1995, \pasj, 47, L5

\bibitem[Gondek et al. 1995]{gon96}
Gondek, D., et al. 1996, \mnras, 282, 646

\bibitem[Hartwick and Schade 1990]{har90}
Hartwick, F.D.A., \& Schade, D. 1990, \araa, 28, 437

\bibitem[Hasinger 1998]{has98}
Hasinger, G. 1998, AN, 319, 37

\bibitem[Lawrence 1991]{law91}
Lawrence, A. 1991, \mnras, 252, 586

\bibitem[McMahon et al. 1992]{mcm92}
McMahon, R.G., Irwin, M.J., \& Hazard, C. 1992, Gemini Issue, 36, 1

%
\bibitem[Schmidt et al. 1998]{sch98}
Schmidt, M., et al.,  1998, A\&A, 329, 495 

\bibitem[Turner and Pounds 1989]{tur89}
Turner, T.J., Pounds, K.A. 1989, \mnras, 240, 833 

\bibitem[Ueda et al. 1998a]{ued98a}
Ueda, Y., et al. 1998a, Nature, 391, 868

\bibitem[Ueda et al. 1998b]{ued98b}
Ueda, Y., et al. 1998b, AN, 319, 47

\bibitem[Ueda et al. 1998c]{ued98c} 
Ueda, Y., et al.  1998c, ApJ, submitted

\bibitem[White et al. 1997]{whi97}
White, R.L., Becker, R.H., Helfand, D.J., \& Gregg, M.D. 1997, \apj, 475, 479

\bibitem[Voges et al. 1998]{vog98}
Voges, W., et al. 1998, in this proceeding

\end{thebibliography}
\end{document}